\documentclass[journal,article,accept,moreauthors,universe]{Definitions/mdpi} 

\firstpage{1} 
\pubvolume{xx}
\issuenum{1}
\articlenumber{5}
\pubyear{2020}
\copyrightyear{2020}
\history{Received: date; Accepted: date; Published: date}

\usepackage{amsmath,amssymb}

\def \apj {ApJ}
\def \apjl {ApJL}

\def \aj {AJ}
\def \aap {A\&A}

\def \mnras {MNRAS}

\def \gray {$\gamma$-ray}
\Title{High-energy and Very High-energy Constraints from Log-parabolic Spectral Models in Narrow-line Seyfert 1 Galaxies}

\Author{Stefano Vercellone $^{1,\dagger,*}$\orcidA{}, Luigi Foschini $^{1}$\orcidC{}, Patrizia Romano $^{1}$\orcidB{}, Markus B{\"o}ttcher $^2$, Catherine Boisson $^3$}

\AuthorNames{Stefano Vercellone, Luigi Foschini, Patrizia Romano, Markus B{\"o}ttcher, Catherine Boisson}

\address{%
$^{1}$ \quad INAF, Osservatorio Astronomico di Brera, Via Emilio Bianchi 46, I-23807 Merate (LC), Italy\\
$^{2}$ \quad Centre for Space Research, North-West University, Potchefstroom, 2531, South Africa\\
$^{3}$ \quad LUTH, Observatoire de Paris, CNRS, Universit\'e  Paris Diderot, PSL Research University Paris, 5 place Jules Janssen, F-92195 Meudon, France\\
}

\corres{Correspondence: stefano.vercellone@inaf.it; Tel.:  +39-02-72320-509}

\firstnote{Current address: INAF/OAB, Via E. Bianchi 46, I-23807 Merate (LC), Italy} 

\abstract{Narrow-line Seyfert 1 galaxies (NLSy1s) are a well-established class of \gray{} sources, showing the presence of a jet like the more common flat-spectrum radio quasars. The evidence of \gray{} emission poses the issue of the location of the \gray{} emitting zone and of the contribution of the $\gamma$-$\gamma$ absorption within the broad-line region (BLR), since such objects have been detected by Fermi-LAT in the MeV-GeV energy range but not by imaging atmospheric Cherenkov telescopes beyond 100\,GeV. We discuss how the spectral properties of three NLSy1s (SBS~0846$+$513, PMN~J0948$+$0022, and PKS~1502$+$036) derived from the Fermi Large Area Telescope Fourth Source Catalog (4FGL) compare with theoretical models based on the observed properties of the BLR. In particular, we focus on the question on how simple power-law spectral models and log-parabolic ones could be disentangled in \gray{} narrow-line Seyfert 1 galaxies by means of current Fermi-LAT or future imaging atmospheric Cherenkov telescopes data. We find that the only possibility for a log-parabolic model to mimic a power-law model in the energy band above $E \sim 100$\,GeV is to have a very small value of the curvature parameter $ \beta \sim 0.05$.
}

\begin{document}
%
%
%
%

\section{Introduction}
The MeV-GeV spectra of jetted active galactic nuclei (AGN) can be fitted by using either power-law or log-parabolic models \cite{2013ApJ...771L...4C,2014ApJ...782...82D,2015ApJ...809..174D}. As the maximum likelihood method to extract spectral parameters in the \gray{} energy band does not allow us to define the goodness of a fit (see, for example, \cite{STAT}), the only way to distinguish between the two models is a clear deviation between data and models visible in the spectra. However, in the Fermi-LAT energy band up to a few tens of GeV, where most of the jetted AGN have been detected, it is not possible to break this degeneracy, unless the curvature of the spectrum is highly significant. Nonetheless, in the latter case, a degeneracy remains between the log-parabola and a broken power-law model. 
We also note that the log-parabola model is more than a simple fitting procedure, but contains an additional physical condition to the jet particle acceleration. As shown in~\cite{2004A&A...413..489M}, this kind of spectra require a dependency on the energy of the acceleration probability.
The problem is important not only in view of locating the region where high-energy gamma rays are emitted, but also to better understand the mechanisms accelerating particles working in a relativistic jet.
The location of the \gray{} emitting region is still an open and very timely issue. There is evidence that in some jetted sources the location of the \gray{} emitting region could reside at different distances from the central black-hole during different flaring episodes of the same source 
as suggested by,  e.g., \cite{2011arXiv1110.4471F} for PKS~1222$+$216.

The large energy band that will be accessible by the Cherenkov Telescope Array (CTA), from $\sim 20$~GeV to hundreds of TeV \cite{2017Msngr.168...21H}, will allow us to break the degeneracy. However, one issue remains open, that is if there are one or more combinations of photon indices and curvature for which the two models hold the degeneracy even in the CTA energy range. Solving this problem is the aim of the present work. 
To complete this task, we extended the simulation work performed to date on a sample of jetted Narrow-Line Seyfert 1 galaxies (NLSy1s) \cite{2018MNRAS.481.5046R,2020arXiv200211737R}, focusing on the possibility to disentangle the different spectral models. 

\section{The Sample}\label{Sec:sample}
Among active galactic nuclei, narrow-line Seyfert 1 galaxies show quite distinctive characteristics. Their definition as a class is due to~\cite{1985ApJ...297..166O}. These sources show narrow permitted emission lines (FWHM $({\rm H} \beta) < 2000$\,km s$^{-1}$), weak ${\rm [O III]}$ lines with a ratio ${\rm [OIII]} / {\rm H} \beta < 3 $, strong optical Iron emission lines (high ${\rm FeII/H} \beta$ ratio), and a relatively low-mass black hole ($10^{6} - 10^{8}$\,M$_{\odot}$) accreting close to the Eddington limit (see~\cite{2011nlsg.confE...2P} and references therein for an historical review of the NLSy1s properties). A first major step forward in the study of this class of sources was given by their detection in the radio band, although only a small fraction of them (4--7\%) shows significant radio emission~\cite{2006AJ....132..531K}. A second breakthrough was their detection in the \gray{} energy band, which revealed the presence of a jet of accelerated particles oriented towards the observer~\cite{2009ApJ...707L.142A}. The latest version of the {\it Fermi Large Area Telescope Fourth Source Catalog} (4FGL, revision v5)~\cite{2019arXiv190210045T} contains four identified and five associated \gray{} NLSy1 galaxies, respectively, but more sources have been found during flaring episodes or dedicated analysis. Interestingly, their spectra can be fit either by a power-law or by a log-parabola model, the latter ones indicating some curvature in their spectra.

A comprehensive sample of \gray{} narrow-line Seyfert 1 galaxies has been discussed by \cite{2018MNRAS.481.5046R}, who collected from literature twenty sources. Among them, three sources seem to be the most promising ones to be investigated at energies above a few tens of GeV, namely SBS~0846$+$513, PMN~J0948$+$0022, and PKS~1502$+$036. We briefly report here their \gray{} discovery literature, while for a detailed description of their high-energy characteristics we refer to \cite{2015A&A...575A..13F, 2016Galax...4...11D, 2019Galax...7...87D} and references therein.
Table~\ref{tab:sample} shows the \gray{} spectral parameters for the log-parabola model as reported in the 4FGL~\cite{2019arXiv190210045T}. The data span eight years of science operations, from 2008 August 4 to 2016 August 2. We note that these parameters were not used in \cite{2018MNRAS.481.5046R}, which in turn generally adopted the power-law model.

\begin{table}[H]
\caption{\gray{} spectral parameters (log-parabola and power-law models) for the three considered sources according to the Fourth Fermi-LAT Catalog \cite{2019arXiv190210045T}. $E_{0}$ is the pivot energy [MeV]; $\alpha$ is the photon index of the log-parabola model; $\beta$ is the curvature; $K_{\rm lp}$ is the normalization of the log-parabola model [$10^{-12}$~ph cm$^{-2}$ s$^{-1}$ MeV$^{-1}$] at the pivot energy; $\Gamma$ is the photon index of the power-law model; $K_{\rm pl}$ is the normalization of the power-law model [$10^{-12}$~ph cm$^{-2}$ s$^{-1}$ MeV$^{-1}$] at the pivot energy. \label{tab:sample}}
\centering
\begin{tabular}{lcccccc}
\toprule
\textbf{Name}	& \textbf{$E_{0}$} & \textbf{$\alpha$} & \textbf{$\beta$} & \textbf{$K_{\rm lp}$} & \textbf{$\Gamma$} & \textbf{$K_{\rm pl}$} \\
\midrule
SBS~0846$+$513	& 624.72	& $2.17\pm0.04$	& $0.08\pm 0.02$ & $8.2\pm 0.3$ 	& $2.27\pm 0.02$ & $7.7\pm 0.2$\\
PMN~J0948$+$0022	& 276.11	& $2.46\pm 0.03$	& $0.16\pm 0.02$ & $163\pm4$	& $2.63\pm 0.02$ & $144\pm 3$\\
PKS~1502$+$036	& 400.94	& $2.48\pm 0.06$	& $0.10\pm 0.03$ & $19\pm1 $	& $2.59\pm 0.04$ & $17.7\pm 0.8$\\
\bottomrule
\end{tabular}
\end{table}

SBS~0846$+$513 ($z=0.585$) was reported as a new \gray{} source in \cite{2011nlsg.confE....F}, as a consequence of a \gray{} flare detected by it Fermi-LAT in 2010, identified by analyzing the first 30 months of data. This source was subsequently confirmed by \cite{2011ATel.3452....1D} and then by \cite{2012MNRAS.426..317D} analyzing the first 40 months of data.

PMN~J0948$+$0022 ($z=0.585$) is the first NLSy1 detected in the \gray{} energy band already during the first months of operations of Fermi-LAT in 2008 \cite{2009ApJ...699..976A}. Two intense \gray{} outbursts were then detected in 2010 July \cite{2011MNRAS.413.1671F} and at the end of 2012 \cite{2015MNRAS.446.2456D}. During the latest outburst, there was also an attempt to detect the source at very high-energy (VHE) with VERITAS at energies $E>200$\,GeV, but without success \cite{2015MNRAS.446.2456D}. We note that VERITAS observations began almost a week after the \gray{} maximum, during the decaying phase of the flare.

PKS~1502$+$036 ($z=0.408$)  was discovered as a \gray{} source in \cite{2009ApJ...707L.142A}, by using the first year of Fermi-LAT data (2008 August 4--2009 August 5).

\section{Broad-line region versus log-parabola spectral model absorption}\label{Sec:BLRvsLP}

\subsection{The log-parabola parameters space in the 4FGL context}\label{Sec:LPparams}
\begin{figure}[H]
\centering
\includegraphics[width=10 cm, angle=90]{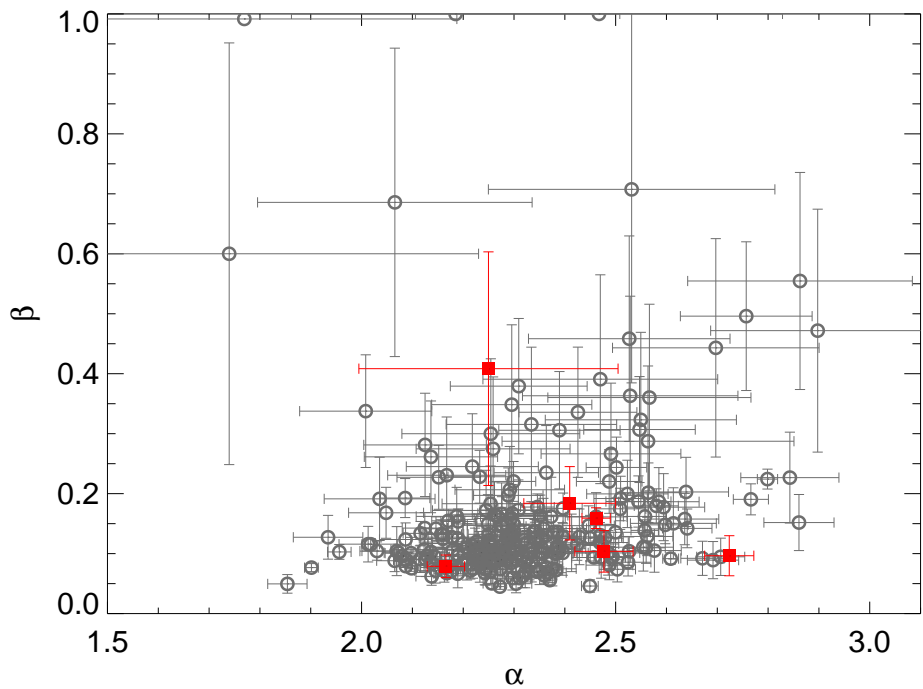}
\caption{Scatter-plot of the two spectral parameters ($\alpha$ and $\beta$) for the different object categories. Open gray circles represent flat-spectrum radio quasars, while filled red squares represent narrow-line Seyfert 1 galaxies, respectively. Data and errors from the 4FGL.
\label{Fig:AlphaBetaScatter} 
}
\end{figure}

Narrow-line Seyfert 1 galaxies have been demonstrated to posses characteristics which resemble those of other extra-galactic jetted sources, in particular flat-spectrum radio quasars (see \cite{2012nsgq.confE..10F} for a review). In order to understand the typical parameter space of the photon index and curvature of the log-parabolic models, we retrieved the values of $(\alpha,\beta)$ for all the quasars and NLSy1s in the 4FGL catalog with the log-parabola indicated as the preferential model. We obtain a sample of 248 quasars and 6 NLSy1s, whose parameters are displayed in Figure~\ref{Fig:AlphaBetaScatter}. It is worth reminding that the curvature parameter in the model adopted by the 4FGL pegs at the value of $1$: therefore, the three sources with $\beta=1$ are a spurious fit
and are not considered in the statistics on $\beta$ reported below.

\begin{figure}[H]
\centering
\includegraphics[width=5.5 cm, angle=90]{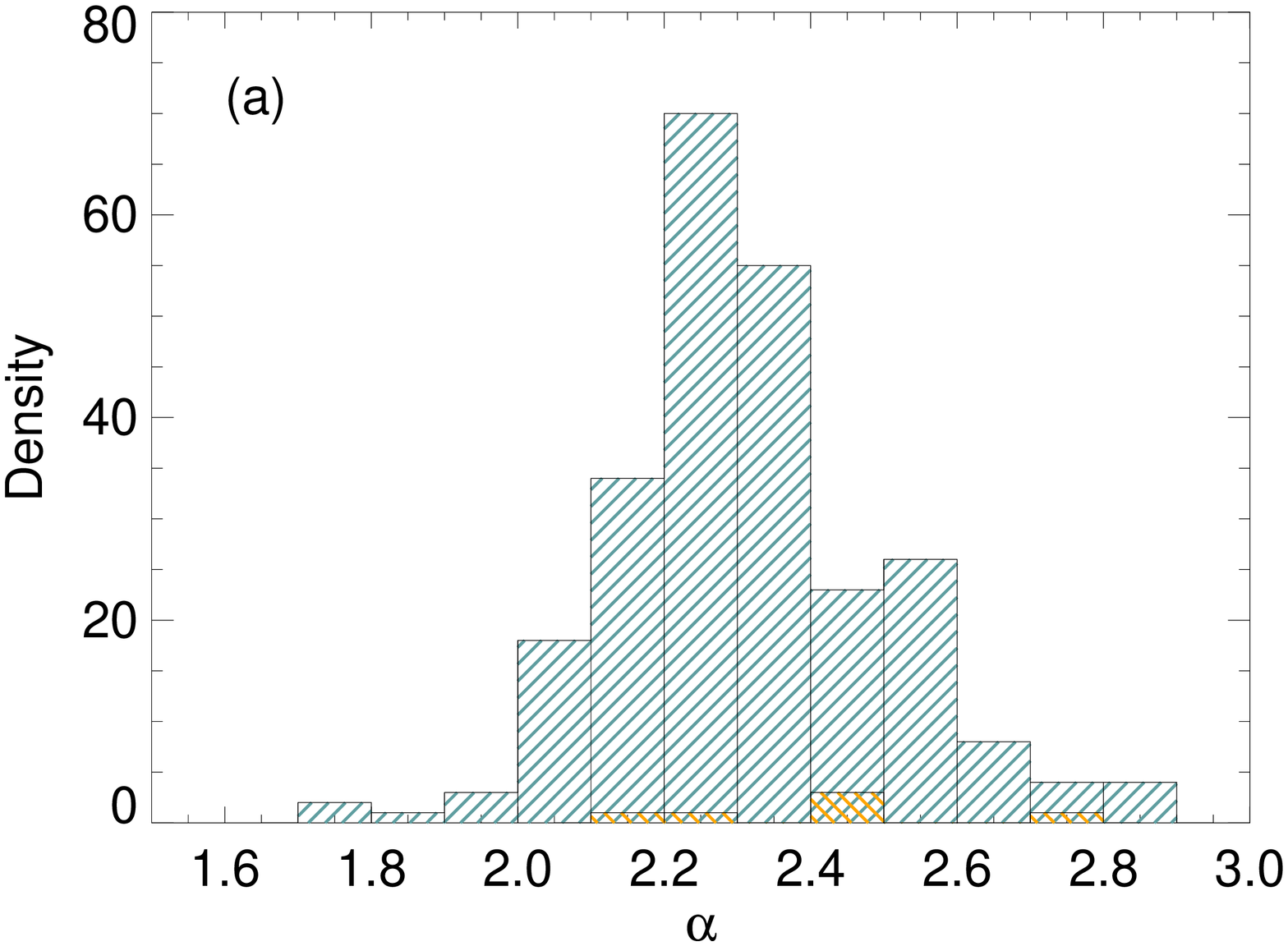}
\includegraphics[width=5.5 cm, angle=90]{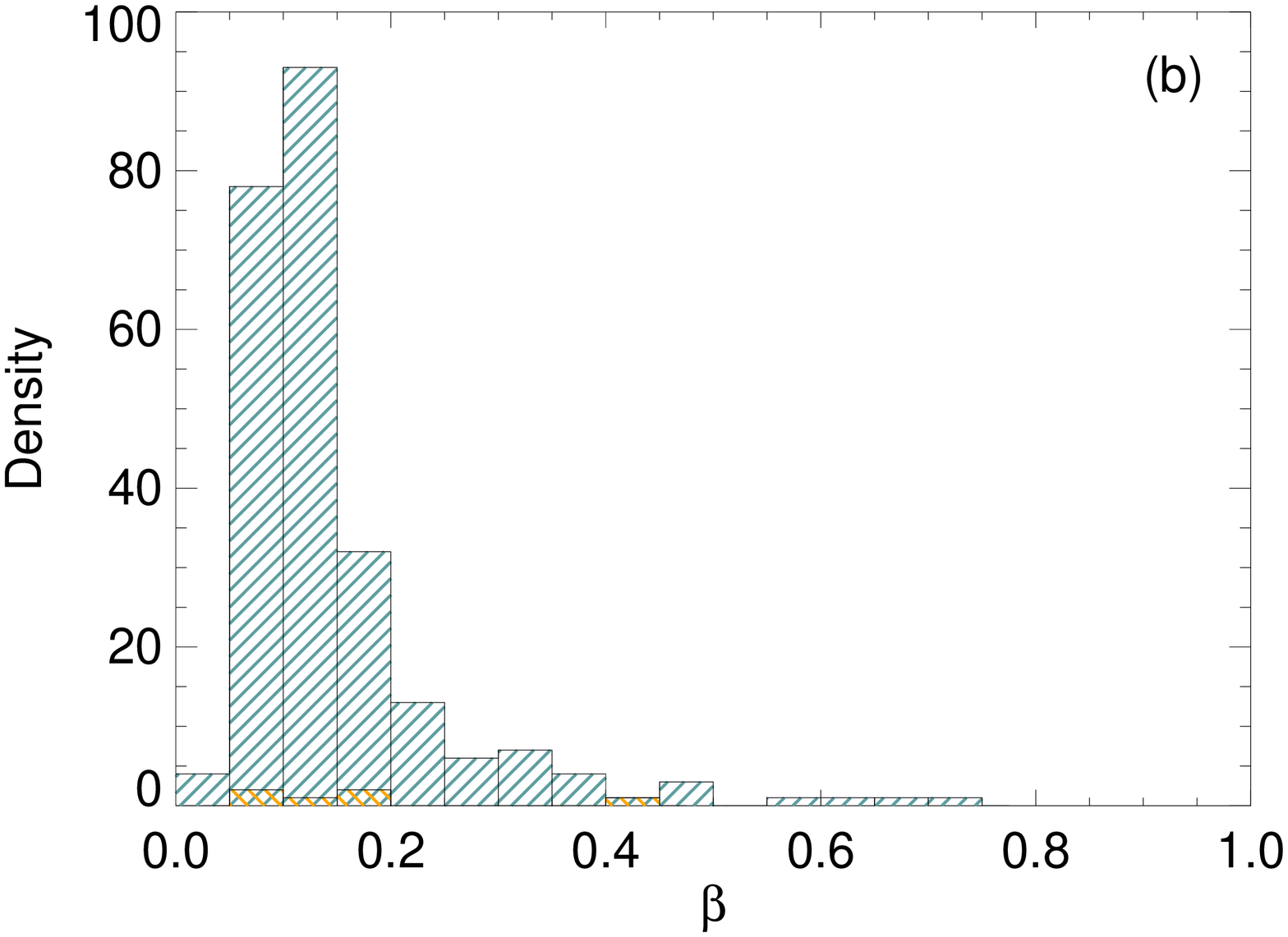}
\caption{(\textbf{a}) Distributions of the $\alpha$ parameter for flat-spectrum radio quasars (blue lines) and narrow-line Seyfert 1 galaxies (orange lines). (\textbf{b}) Distributions of the $\beta$ parameter for flat-spectrum radio quasars (blue lines) and narrow-line Seyfert 1 galaxies (orange lines). 
\label{Fig:AlphaBetaDistro} }
\end{figure}

We can now consider the distributions of the two spectral parameters. These distributions will allow us to put initial constraints on $\alpha$ and $\beta$. Figure~\ref{Fig:AlphaBetaDistro} shows the distributions of both the $\alpha$ and $\beta$ parameters for the whole sample of flat-spectrum radio quasar (blue lines) and narrow-line Seyfert 1 galaxies (orange lines), respectively. We obtain $\alpha = 2.32 \pm 0.19$,  while $\beta = 0.15\pm0.10$, with $1.74 \le \alpha \le 2.90$ and $0.05 \le \beta \le 0.71$. We should note, however, that while the distribution on the $\alpha$ parameter has a low skewness value ($sk = 0.33$), the distribution on the $\beta$ parameter has a high skewness value ($sk = 2.73$). A large fraction, $84$\%, of $\beta$ values are equal or lower than 0.2. This could be a hint of a non robust log-parabolic fit of the spectrum of those sources in Figure~\ref{Fig:AlphaBetaScatter} with $\beta \ge (0.5-0.6)$ whose error-bars are particularly large. The small average value of the curvature also explains the degeneracy with the power-law model in the Fermi-LAT energy band.

\subsection{Application to the three NLSy1s}\label{Sec:LPNLSY1}
As reported in Section~\ref{Sec:sample}, we chose SBS~0846$+$513, PMN~J0948$+$0022, and PKS~1502$+$036 because, in principle, during flaring events they could be detected by the next generation Cherenkov Telescope Array~\cite{2017Msngr.168...21H}, as discussed in~\cite{2020arXiv200211737R}, where the authors considered a detailed treatment of $\gamma-\gamma$ absorption in the radiation fields of the BLR as a function of the location of the $\gamma$-ray emission region with parameters inferred from observational constraints~\cite{2016ApJ...821..102B}. In their Figure~7, the authors compared the log-parabola models for the three sources (whose spectral parameters where derived from the 4FGL), renormalised to the integrated 100\,MeV to 100\,GeV flux as extrapolated by adopting the power-law models detailed in their Table~1, with respect to the expected spectra due to the BLR internal $\gamma-\gamma$ absorption. The comparison shows that the log-parabola models are distinct from the BLR internal $\gamma-\gamma$ absorption.
However, it is necessary to take into account that there can be spectral changes during outbursts and some quasars have shown radical spectral changes, suggesting a change of the location of the \gray{} emission (e. g. \cite{2011arXiv1110.4471F,2013MNRAS.432L..66G}). Therefore, it is worth exploring if there is any realistic combination of $(\alpha,\beta)$ and fluxes for which CTA cannot break the degeneracy with the power-law model.

We simulated a grid of $ 9 \times 9 $ values for $\alpha$ and $\beta$, ($\alpha \in [1.5,2.9]$; $\beta \in [0.05,0.2]$). We considered both the BLR contribution to absorption and the attenuation due to the extra-galactic background light (EBL, calculated by adopting the model of~\cite{2011MNRAS.410.2556D}).

\begin{figure}[H]
\centering
\vspace{-6truecm}
\includegraphics[width=16 cm, angle=0]{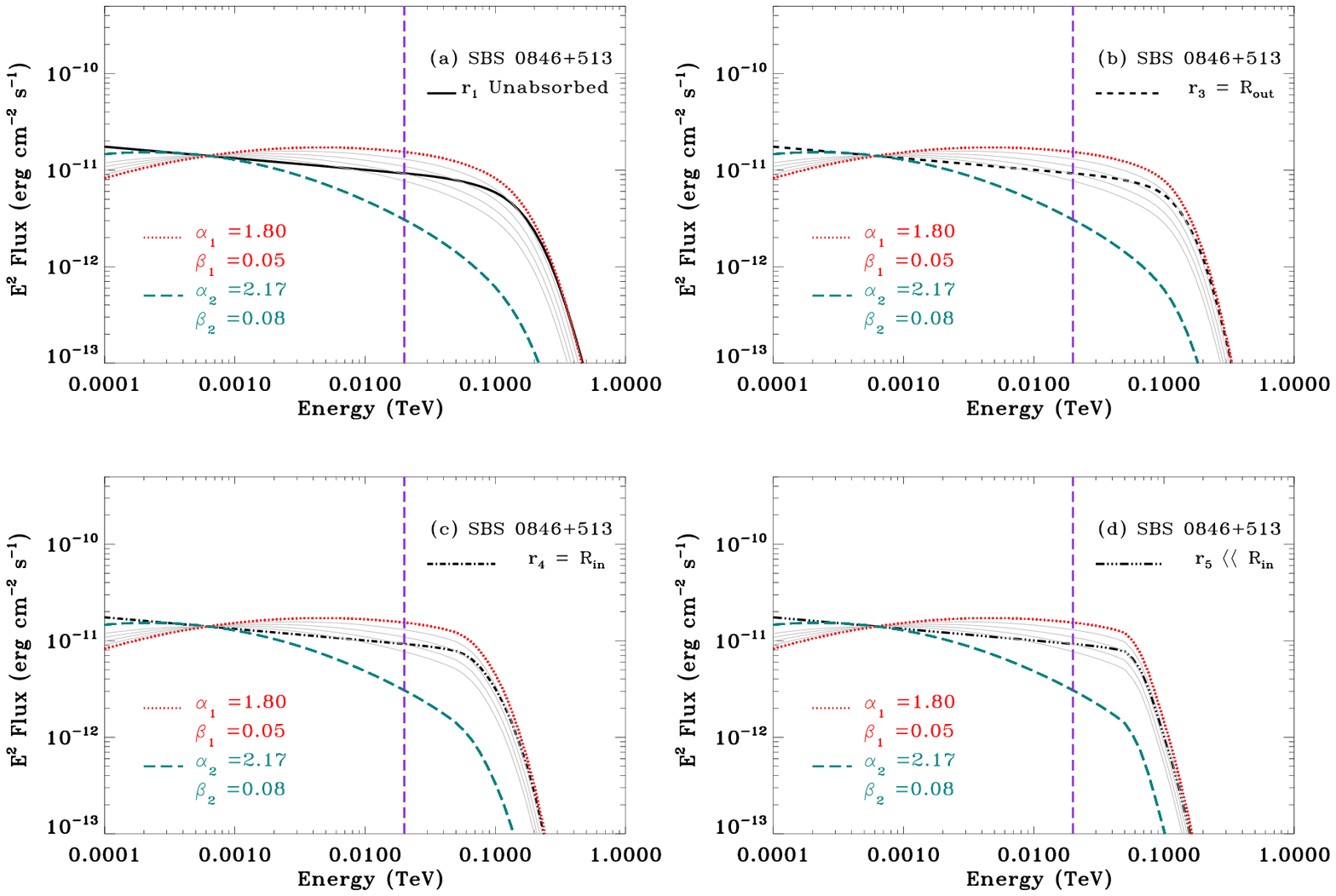}
\vspace{-7truecm}
\caption{Spectral energy distribution in the energy range 100\,MeV -- 1\,TeV for SBS~0846$+$513. (\textbf{a}) $R_{\rm em} = r_{1} \gg R_{\rm BLR}$; (\textbf{b}) $R_{\rm em} = r_{3} = R_{\rm out}$; (\textbf{c}) $R_{\rm em} = r_{4} = R_{\rm in}$; (\textbf{d}) $R_{\rm em} = r_{5} \ll R_{\rm in}$. The vertical purple dashed line at $E = 20$\,GeV represents the CTA energy lower bound.} \label{Fig:SedBoettcherLogPar0846}
\end{figure}

\begin{figure}[H]
\centering
\vspace{-6truecm}
\includegraphics[width=16 cm, angle=0]{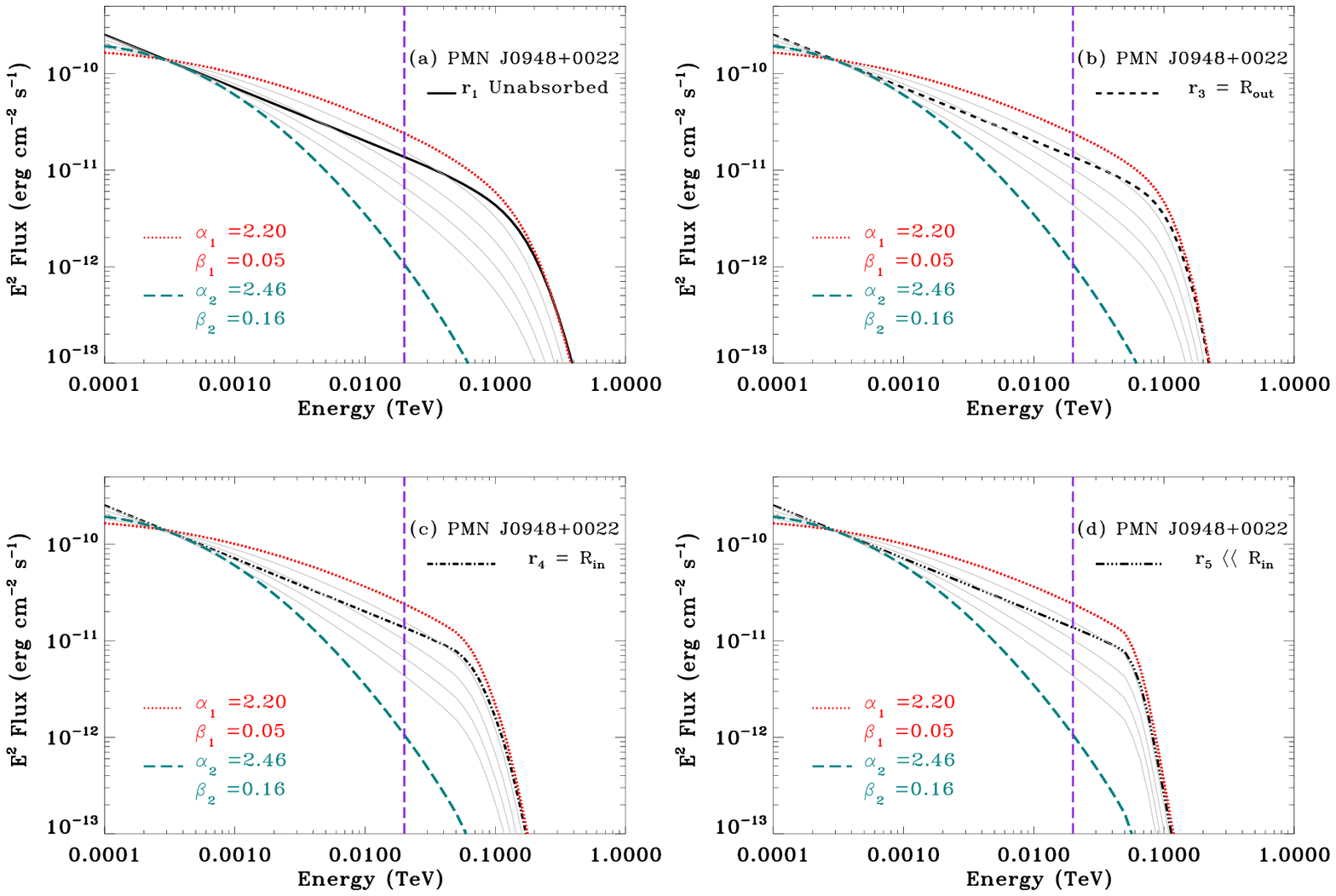}
\vspace{-7truecm}
\caption{Spectral energy distribution in the energy range 100\,MeV -- 1\,TeV for PMN~0948$+$0022. (\textbf{a}) $R_{\rm em} = r_{1} \gg R_{\rm BLR}$; (\textbf{b}) $R_{\rm em} = r_{3} = R_{\rm out}$; (\textbf{c}) $R_{\rm em} = r_{4} = R_{\rm in}$; (\textbf{d}) $R_{\rm em} = r_{5} \ll R_{\rm in}$. The vertical purple dashed line at $E = 20$\,GeV represents the CTA energy lower bound.} \label{Fig:SedBoettcherLogPar0948}
\end{figure}

\begin{figure}[H]
\centering
\vspace{-6truecm}
\includegraphics[width=16 cm, angle=0]{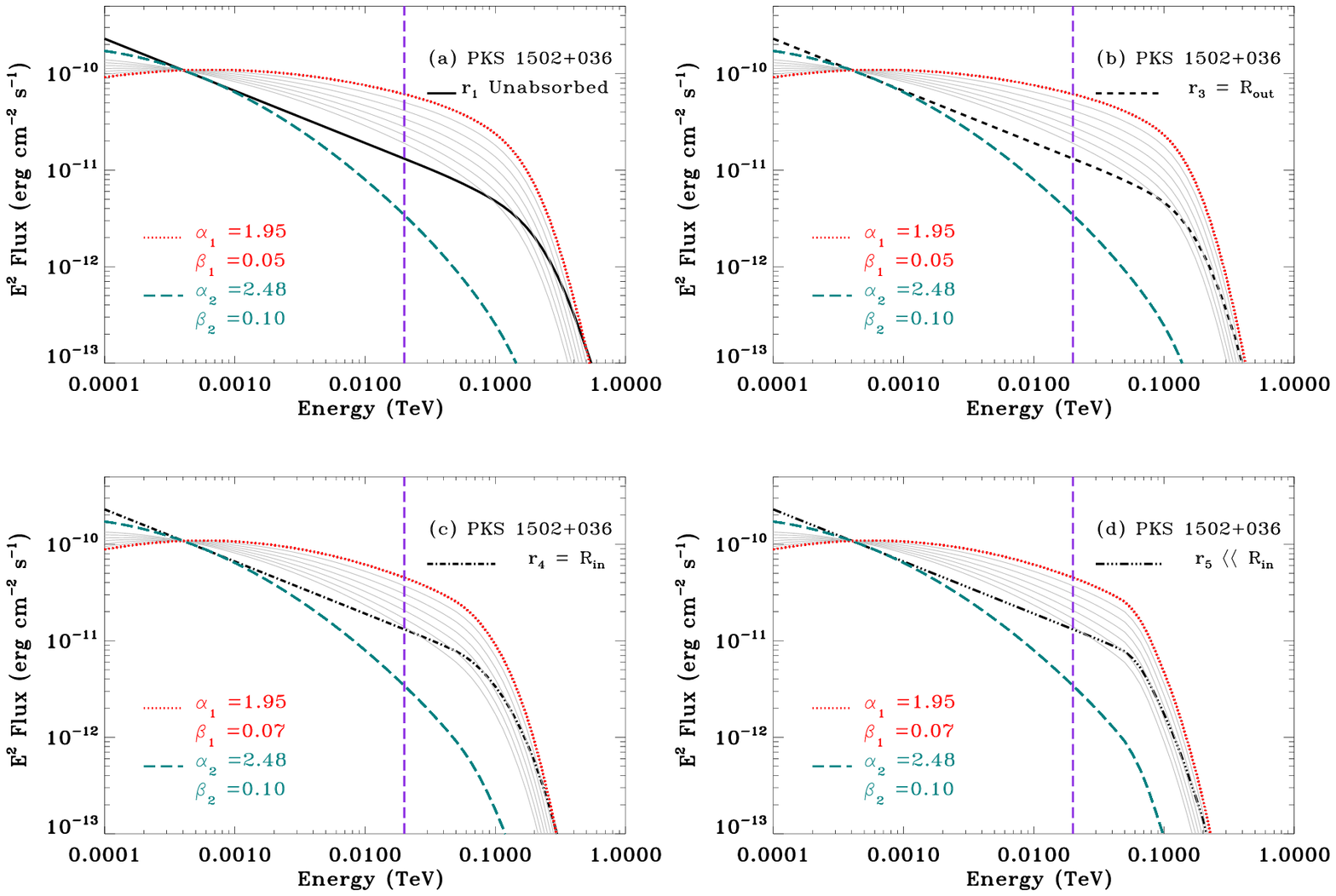}
\vspace{-7truecm}
\caption{Spectral energy distribution in the energy range 100\,MeV -- 1\,TeV for PKS~1502$+$036. (\textbf{a}) $R_{\rm em} = r_{1} \gg R_{\rm BLR}$; (\textbf{b}) $R_{\rm em} = r_{3} = R_{\rm out}$; (\textbf{c}) $R_{\rm em} = r_{4} = R_{\rm in}$; (\textbf{d}) $R_{\rm em} = r_{5} \ll R_{\rm in}$. The vertical purple dashed line at $E = 20$\,GeV represents the CTA energy lower bound.} \label{Fig:SedBoettcherLogPar1502}
\end{figure}

Figure~\ref{Fig:SedBoettcherLogPar0846},~\ref{Fig:SedBoettcherLogPar0948}, and~\ref{Fig:SedBoettcherLogPar1502} show the spectral energy distributions (SEDs) for SBS~0846$+$513, PMN~0948$+$0022, and PKS~1502$+$036, respectively, in the range 100\,MeV -- 1\,TeV. 
All SEDs have been computed fixing the pivot energy, ${\rm E}_{0}$, to the one extracted from the 4FGL for each source and allowing the normalization factor, ${\rm K}_{0}$, to vary (we are analysing possible outbursts) in order to intercept the black line curves values at ${\rm E}_{0}$. We note that this is not a constraint on the flux computed on the full Fermi-LAT energy band (0.1--300)\,GeV, but just at the energy ${\rm E}_{0}$. The black lines were calculated in~\cite{2020arXiv200211737R} starting from a simple power-law model and represent the SEDs obtained by means of a convolution of the internal BLR $\gamma-\gamma$ absorption with the EBL attenuation. The different lines refer to different radii at which the \gray{} emission location, $R_{\rm em}$, is placed with respect to the BLR boundaries, from least absorbed to most absorbed, following the enumeration adopted in~\cite{2020arXiv200211737R}: 
\begin{itemize}
    \item solid line $R_{\rm em} = r_{1} \gg R_{\rm BLR}$, 
    \item dashed lines $R_{\rm em} = r_{3} = R_{\rm out}$ (outer BLR radius),
    \item dot-dashed lines $R_{\rm em} = r_{4} = R_{\rm in}$ (inner BLR radius),
    \item and dashed-triple-dotted lines $R_{\rm em} = r_{5} \ll R_{\rm BLR}$, respectively. 
\end{itemize}

The red dotted lines represent the log-parabola models whose slopes are adjacent to the portion of the SED at VHE ($E >$ a few hundreds of GeV, black solid lines) which were obtained by means of a simple power-law model, as discussed in~\cite{2020arXiv200211737R}.
In Figure~\ref{Fig:SedBoettcherLogPar1502}, panels (c) and (d), the value of $\beta = 0.07$ provides a better agreement with respect to the black lines at energies above a few hundred of GeV.
The thin gray lines represent the various SEDs, obtained by means of a log-parabolic model, with ${\rm E}_{0}$ and ${\rm K}_{0}$ as described above, fixing $\beta$ to the value obtained from the red dotted fit, and allowing $\alpha$ to vary with steps of 0.05.
The green long-dashed lines represent the log-parabola models whose spectral parameters were derived from the 4FGL, except for the normalization factor which has been chosen as described above.

\begin{figure}[H]
\centering
\includegraphics[width=5.5 cm, angle=90]{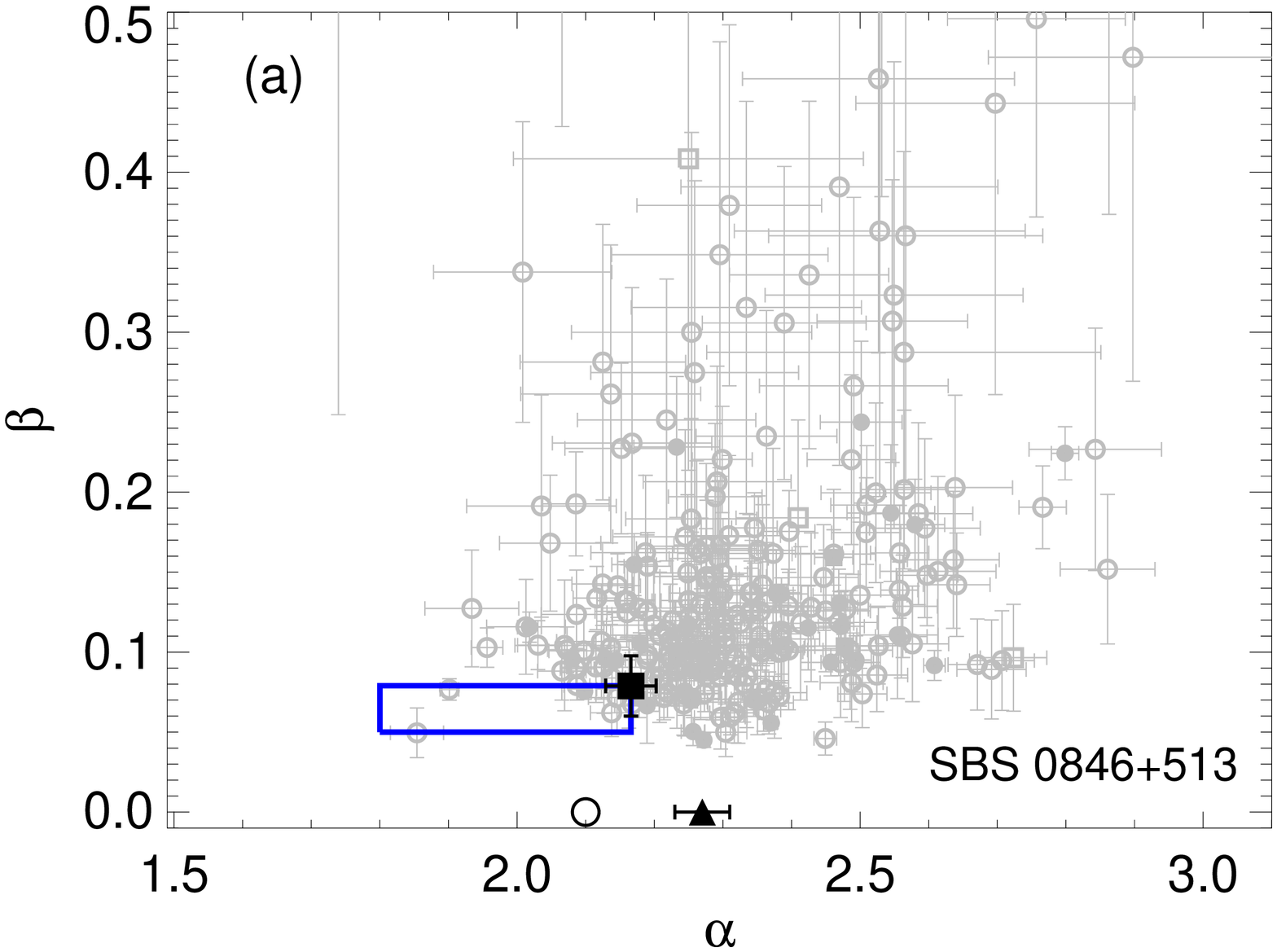}
\includegraphics[width=5.5 cm, angle=90]{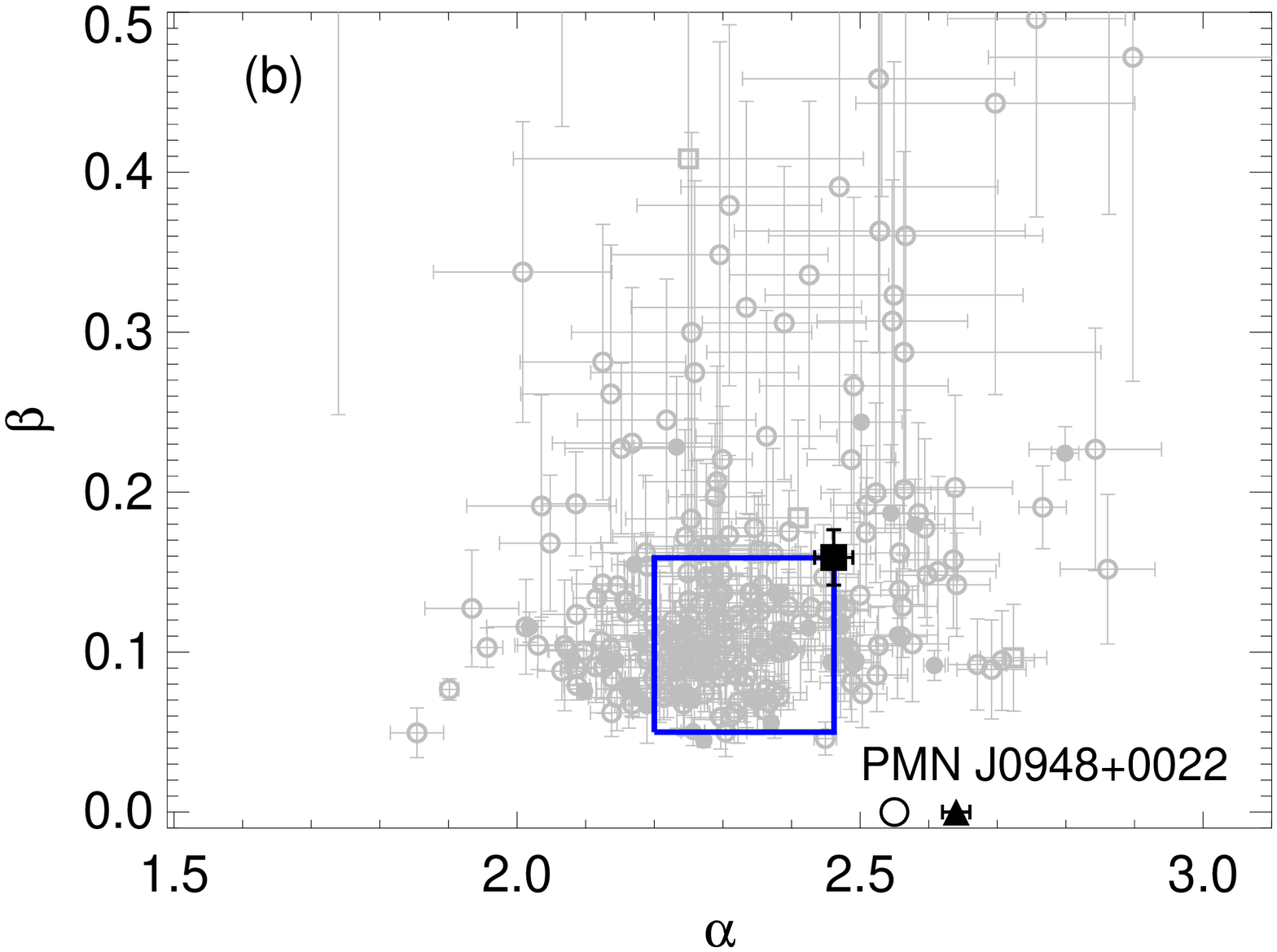}\\
\includegraphics[width=5.5 cm, angle=90]{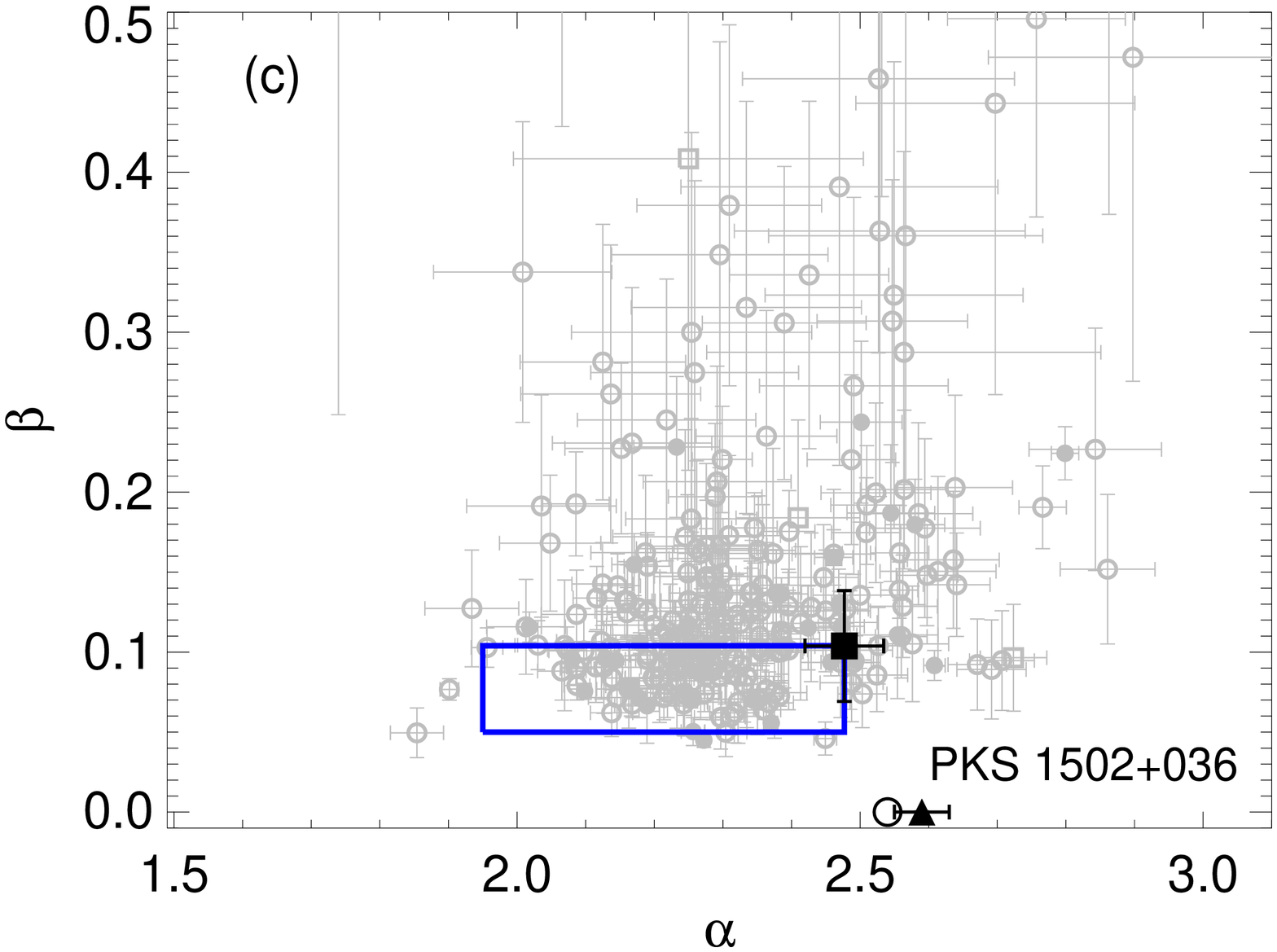}\\
\caption{Scatter-plot of the two spectral parameters ($\alpha$ and $\beta$) for the different object categories (see Figure~\ref{Fig:AlphaBetaScatter} for the description of the different symbols). The black filled squares represent the 4FGL values, while the black filled triangles represent the value of the photon index $\Gamma$ for a power-law fit in 4FGL. The black open circles represent the value of the photon index $\Gamma$ for the different sources in the high \gray{} state as reported in~\cite{2020arXiv200211737R}. Blue rectangles represent the intervals, in $\alpha$ and $\beta$, defined by the red and green dashed curves in Figure~\ref{Fig:SedBoettcherLogPar0846}. (\textbf{a}) SBS~0846$+$513. (\textbf{b}) PMN~0948$+$0022. (\textbf{c}) PKS~1502$+$036. \label{Fig:AlphaBetaBoxes}}
\end{figure}

Figure~\ref{Fig:AlphaBetaBoxes} shows the scatter-plot of the two spectral parameters ($\alpha$ and $\beta$) for the different object categories (see Figure~\ref{Fig:AlphaBetaScatter} for the description of the different symbols), where all the sources have been reported in gray except the three discussed in details in this work. The black filled squares represent the 4FGL log-parabola values as in Figure~\ref{Fig:AlphaBetaScatter}, while the black filled triangles represent the value of the photon index $\Gamma$ for a power-law fit in 4FGL. The black open circles represent the value of the photon index $\Gamma$ for the different sources in high \gray{} state as reported in~\cite{2020arXiv200211737R}. We also added open squares which represent the intervals, in $\alpha$ and $\beta$, defined by the red and green dashed curves in Figure~\ref{Fig:SedBoettcherLogPar0846}, ~\ref{Fig:SedBoettcherLogPar0948}, and~\ref{Fig:SedBoettcherLogPar1502}. Panels (\textbf{a}), (\textbf{b}), and (\textbf{c}) represent SBS~0846$+$513, PMN~0948$+$0022, and PKS~1502$+$036, respectively.

\section{Discussion}\label{Discussion}
In this work we addressed the question on how simple power-law spectral models and log-parabolic ones could be disentangled in \gray{} narrow-line Seyfert 1 galaxies by means of current Fermi-LAT or future CTA data. In particular, we simulated different sets of log-parabolic spectra -whose boundaries where extracted from the 4FGL- extending from a few hundred of MeV up to TeV energy band.
Figures~\ref{Fig:SedBoettcherLogPar0846},~\ref{Fig:SedBoettcherLogPar0948}, and~\ref{Fig:SedBoettcherLogPar1502} show that there are sets of log-parabola models --i.e., different combinations of ($\alpha$,$\beta$)-- which mimic the behaviour of a power-law at energies of $E > 200-300$\,GeV.
The simulations show that the only possibility for a log-parabolic model to mimic a power-law model in the CTA energy band is to have a very small value of the curvature parameter $\beta \sim 0.05$. To have an idea of the improvement with respect to Fermi-LAT alone, it is possible to compare this result with what has been found in~\cite{2012A&A...548A.106F} by using four-year data of PMN~0948$+$0022: in that case, the data did not allow the authors to disentangle between a fit of the overall spectrum with a power-law and a log-parabola models, with a curvature parameter $\beta \sim 0.3$. Since, as shown in Fig.~\ref{Fig:AlphaBetaScatter}, most of the quasars have values of $\beta \lesssim 0.3$, this means that CTA will allow us to break the spectral degeneracy for a large number of sources. Quasars with $\beta \lesssim 0.05$, for which CTA will not be able to distinguish between different models, represent a small population in the 4FGL catalogue. If we impose $\alpha \le 2.5$ and $\beta \le 0.05$, the following flat-spectrum radio quasars satisfy this condition: 4FGL~J0043.8$+$3425 ($\alpha = 1.854 \pm 0.039$, $\beta = 0.050 \pm 0.016$), 4FGL~J0112.8$+$3208 ($\alpha = 2.304 \pm 0.031$, $\beta = 0.050 \pm 0.015$), 4FGL~J2143.5$+$1743 ($\alpha = 2.449 \pm 0.016$, $\beta = 0.046 \pm 0.010$), and 4FGL~J1224.9$+$2122 ($\alpha = 2.272 \pm 0.009$, $\beta = 0.045 \pm 0.004$). We also note that the SBS~0846$+$513 spectrum fit in the 4FGL yield $\beta = 0.079$, very close to this limiting value. 

\begin{table}[H]
\caption{\gray{} spectral parameters and fluxes (log-parabola models) for the three considered sources. {\bf Model} refers to the red or green models in Figures~\ref{Fig:SedBoettcherLogPar0846}, \ref{Fig:SedBoettcherLogPar0948}, and~\ref{Fig:SedBoettcherLogPar1502}. $F_{\rm (140-200)\,GeV}^{\rm mod}$ and $F_{\rm (200-280)\,GeV}^{\rm mod}$ are the fluxes in units of [$10^{-13}$~erg cm$^{-2}$ s$^{-1}$], while $F_{\rm (140-200)\,GeV}^{\rm sim}$ and $F_{\rm (200-280)\,GeV}^{\rm sim}$ are those reported in Tables~4, 5, and 7 of~\cite{2020arXiv200211737R}, same units.\\
Notes: $^{(1)}$: curve with $\alpha=1.95$ and $\beta=0.05$. $^{(2)}$: curve with $\alpha=1.95$ and $\beta=0.07$.} \label{tab:results}
\centering
\begin{tabular}{lcccccc}
\toprule
\textbf{Name}	& \textbf{Model}  & \textbf{$r_{\rm n}$} & \textbf{$F_{\rm (140-200)\,GeV}^{\rm mod}$} & \textbf{$F_{\rm (200-280)\,GeV}^{\rm mod}$} & \textbf{$F_{\rm (140-200)\,GeV}^{\rm sim}$} & \textbf{$F_{\rm (200-280)\,GeV}^{\rm sim}$} \\
\midrule
SBS~0846$+$513	& Red & -	& 32.3 & 26.8 & - & -\\
	& - & $r_{1}$	& - & - & 32.3 & 14.9 \\
	& Green & -	& 1.72 & 1.11 & - & -\\
PMN~J0948$+$0022 & Red & -	& 18.3 & 12.8 & - & -\\
	& - & $r_{1}$	& - & - & 19.9 & 7.8 \\
	& Green & -	& 0.04 & 0.02 & - & -\\
PKS~1502$+$036 & Red$^{(1)}$ & -	& 85.1 & 65.9 & - & -\\
    & Red$^{(2)}$ & -	& 41.3 & 29.3 & - & -\\
	& - & $r_{1}$	& - & - & 26.8 & 14.6 \\
	& Green & -	& 0.51 & 0.26 & - & -\\
\bottomrule
\end{tabular}
\end{table}

We see that integrating in the full Fermi-LAT energy band ($0.1 \le E_{\rm GeV} \le 300$) the difference in flux between the simple power-law models described in~\cite{2020arXiv200211737R} and the log-parabolic models described here are on the order of the 10--30\,\%. On the other hand, differences in fluxes could be conspicuous in more restricted energy bands, (140-200)\,GeV and (200-280)\,GeV, typical of VHE telescopes arrays.
Table~\ref{tab:results} shows the fluxes in two different energy bands, (140-200)\,GeV and (200-280)\,GeV. We compare the ``red'' and ``green'' models with respect to the un-absorbed, $R_{\rm em} = r_{1}$, one. We clearly see that, while the ``red'' and $r_{1}$ models yield similar fluxes (within a factor of $\sim 2-4$), they could be more than one order of magnitude higher with respect to the ``green'' one.
%

\section{Conclusions}\label{Conclusions}
The forthcoming Cherenkov Telescope Array will play a relevant role in the possible detection of NLSy1s during flares. This will allow us to perform almost simultaneous multi-wavelength observations, covering a wide energy band from radio frequencies up to several hundreds of GeV, which will be extremely important to investigate the spectral properties of these sources. We have seen that Fermi-LAT data alone may not be conclusive in disentangling between a power-law and a log-parabolic model of the NLSy1s spectra. 
We can conclude that, although the BLR absorption tends to strongly affect any spectra, a log-parabola with significant curvature can be distinguished from a power-law model in the CTA energy band.
%

%



\vspace{6pt} 



\authorcontributions{Conceptualization, L.F., S.V., P.R.; Writing--Original Draft Preparation, S.V.; Writing--Review and Editing, S.V, L.F., P.R., M.B., C.B.}

\funding{S.V., L.F., and P.R. acknowledge financial contribution from the agreement ASI--INAF n. 2017-14-H.0.\\
The work of M.B. is supported through the South African Research Chairs Initiative (SARChI) of the Department of Science and Innovation and the National Research Foundation\footnote{Any opinion, finding and conclusion or recommendation expressed in this material is that of the authors, and the NRF does not accept any liability in this regard.} of South Africa.
}

\acknowledgments{We thank the Referees for their prompt and constructive comments.}

\conflictsofinterest{The authors declare no conflict of interest.}


\reftitle{References}

\end{document}